\documentclass[11pt]{article}
\usepackage{moriond,epsfig}

\def\Journal#1#2#3#4{{#1} {\bf #2}, #3 (#4)}

\def\NPB{{\em Nucl. Phys.} B}
\def\PLB{{\em Phys. Lett.}  B}
\def\PRL{\em Phys. Rev. Lett.}
\def\PRD{{\em Phys. Rev.} D}
\def\RMP{\em Rev. Mod. Phys.}

\def\JHEP{\em J.\ High Energy Phys.}

\def\be{\begin{equation}}
\def\ee{\end{equation}}
\def\bea{\begin{eqnarray}}
\def\eea{\end{eqnarray}}

\def\MET{{\mbox{$E\kern-0.57em\raise0.19ex\hbox{/}_{T}~$}}}
\def\METnoSpace{{\mbox{$E\kern-0.57em\raise0.19ex\hbox{/}_{T}$}}}

\begin{document}

\vspace*{4cm}

\title{SEARCHES FOR NEW PHYSICS AT THE TEVATRON}

\author{ M. Jaffr\'e\\on behalf of the CDF and D0 collaborations}

\address{LAL, Universit\'e Paris-Sud, CNRS/IN2P3,\\
91898 Orsay Cedex, France}

\maketitle\abstracts{
The Tevatron collider has provided the CDF and D0 experiences with large datasets
as input to a rich program of searches for physics beyond the standard model.
The results presented here are a partial survey of recent searches conducted by the two collaborations  using up to 6~fb$^{-1}$ of data.
}

\section{Introduction}

The standard model (SM) of particles, despite its remarkable description of experimental data at the elementary particle level,
has some deficiencies to explain what is observed in the universe~: lack of anti-matter, existence of dark matter,\ldots
Working at the energy frontier, as was the case at the Tevatron for so many years, gives experimentalists the hope to discover
new non-SM particles which would indicate some direction to follow at explaining these SM deficiencies.

Over the years, the CDF and D0 experiments have gained experience in the detector responses to all particle types.
It allows to look at a large number of different final states searching for deviations from the SM expectations.
As the knowlegde of detector particle responses becomes more accurate, the complexity of final states can increase.  


For a given final state signature, the non-observation of deviations from the SM prediction allows to constrain several models at once.

\section{Dielectron, dimuon or diphoton resonance searches}
Experimentally, in a hadronic environment, dielectron, dimuon or diphoton final states are easy to identify.
A lot of extensions to the SM predicts the existence of new particles which could be observed as narrow resonances
decaying into a pair of leptons or photons.
Among those are new spin-1 gauge boson~\cite{LANGACKER} or spin-2 Randall-Sundrum (RS) graviton~\cite{RS}.

\begin{figure}[htb]
\centering 
 \begin{minipage}[t]{0.47\linewidth}
    \includegraphics[width=7.7cm]{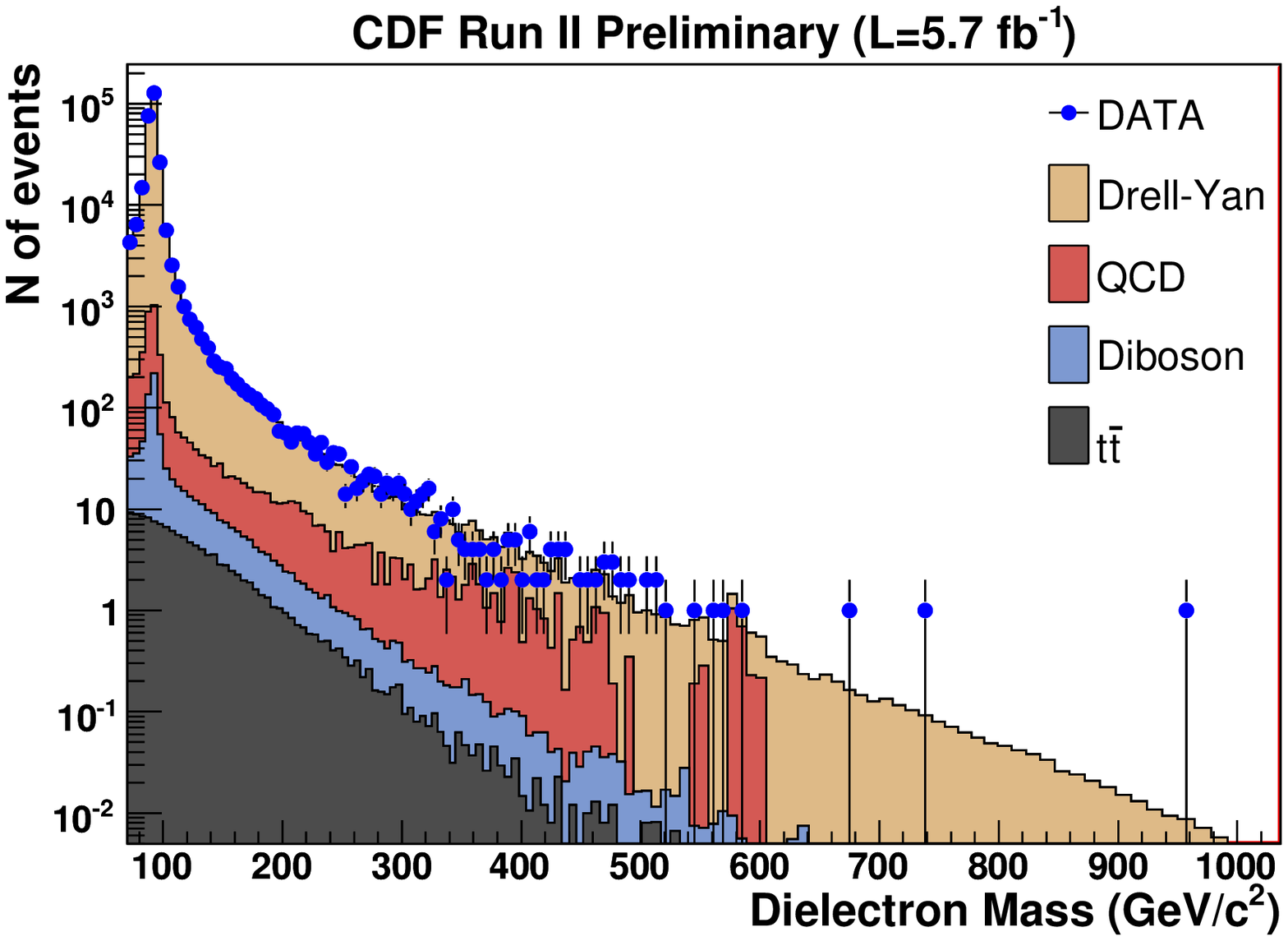}
    \caption{\label{fig:cdf-mee}
    Invariant dielectron mass distribution compared to the expected backgrounds in CDF data.}
  \end{minipage}
  \hspace{5mm}
  \begin{minipage}[t]{0.47\linewidth}
    \includegraphics[width=7.7cm]{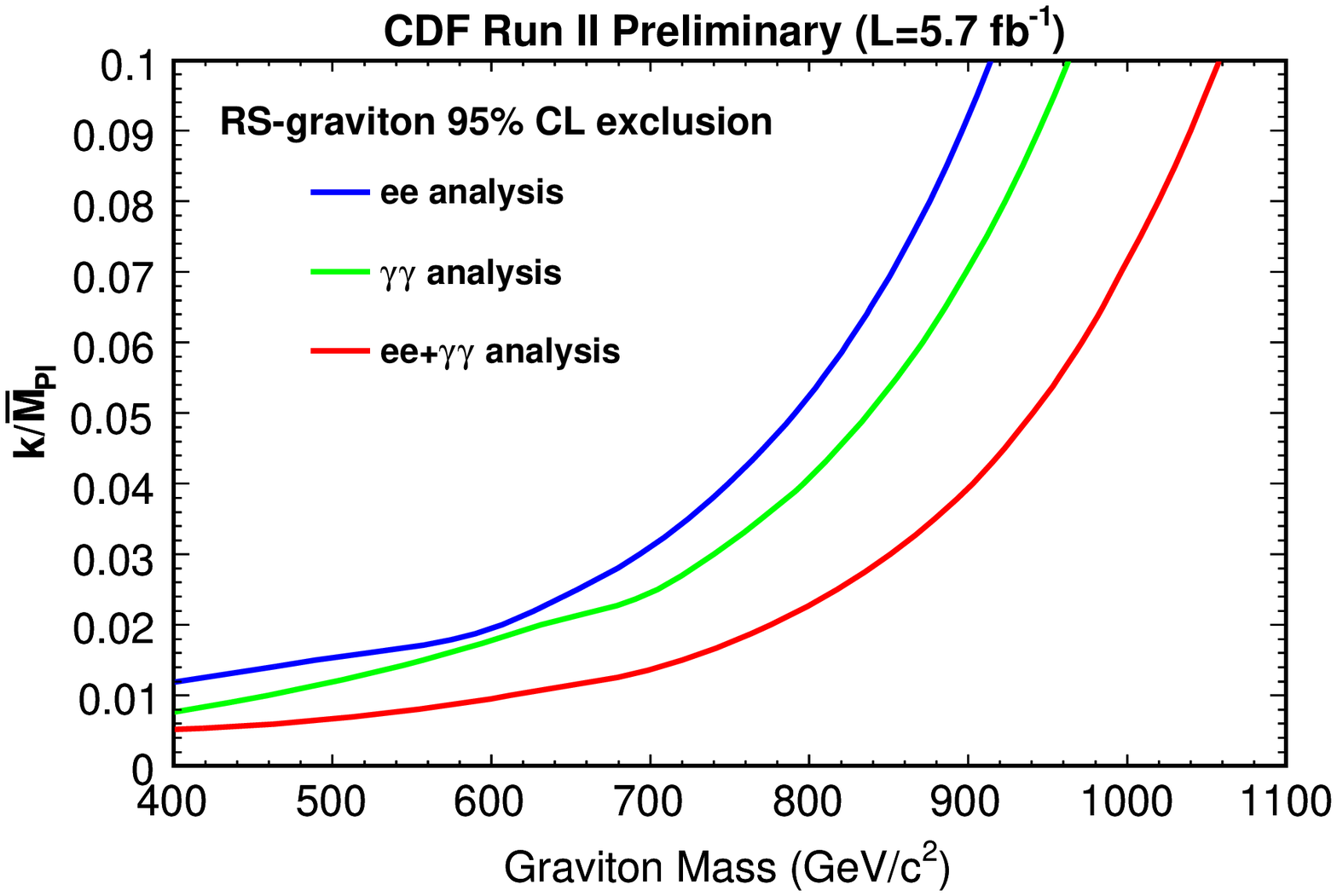}
    \caption{\label{fig:cdf-graviton-limits}
    Excluded domains in the RS parameter space ($k/\overline{M}_{Pl}$, graviton mass) obtained in the dielectron analysis,
the diphoton analysis and their combination}
  \end{minipage}
\end{figure}

The observed dilepton mass distributions observed in large datasets (around 5~fb$^{-1}$ or more) by CDF~\cite{cdf-ee+gg},\cite{cdf-mumu} and
D0~\cite{d0-ee} are in agreement with the SM expectations.
The largest invariant dielectron mass observed by CDF is 960~GeV(Fig.~\ref{fig:cdf-mee}).
No new resonance is observed above the $Z$ gauge boson mass. This allows both experiments to set 95\% C.L. cross section limits which
can be compared to model predictions; as an example, the benchmark SM-like $Z'$ is now excluded  for masses below 1 TeV.

The same spectra can be used to extract limits for the graviton $G$ mass; higher limits can be achieved with the diphoton spectra~\cite{cdf-gg}
as the $G$ branching fraction into two photons is twice that into a lepton pair.
 But, the best limits are obtained by combining dielectron and diphoton analyses\cite{d0-ee+gg,cdf-ee+gg}.
Fig.~\ref{fig:cdf-graviton-limits} shows the excluded domain, obtained by CDF, in the two parameter space of the RS model,
 $k/\overline{M}_{Pl}$, the ratio of curvature of the extra dimension to reduced Planck scale, and the $G$ mass.

\vspace{-5mm}
\section{Diboson resonance search}

\begin{figure}[htb]
  \centering
  \begin{picture}(160,45)
  \put(0,0){\hbox{
 \includegraphics[width=8.0cm]{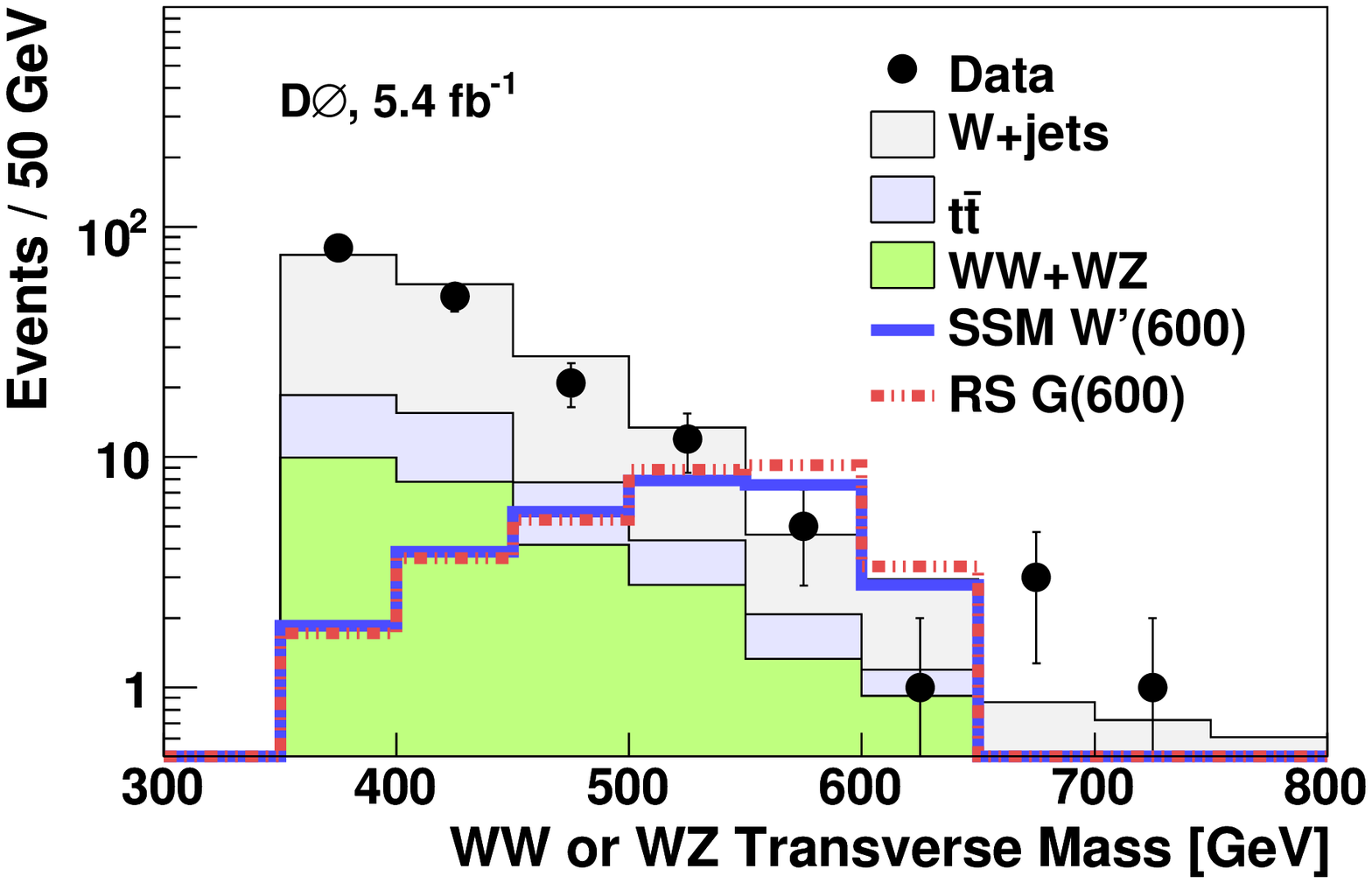}
 \includegraphics[width=8.0cm]{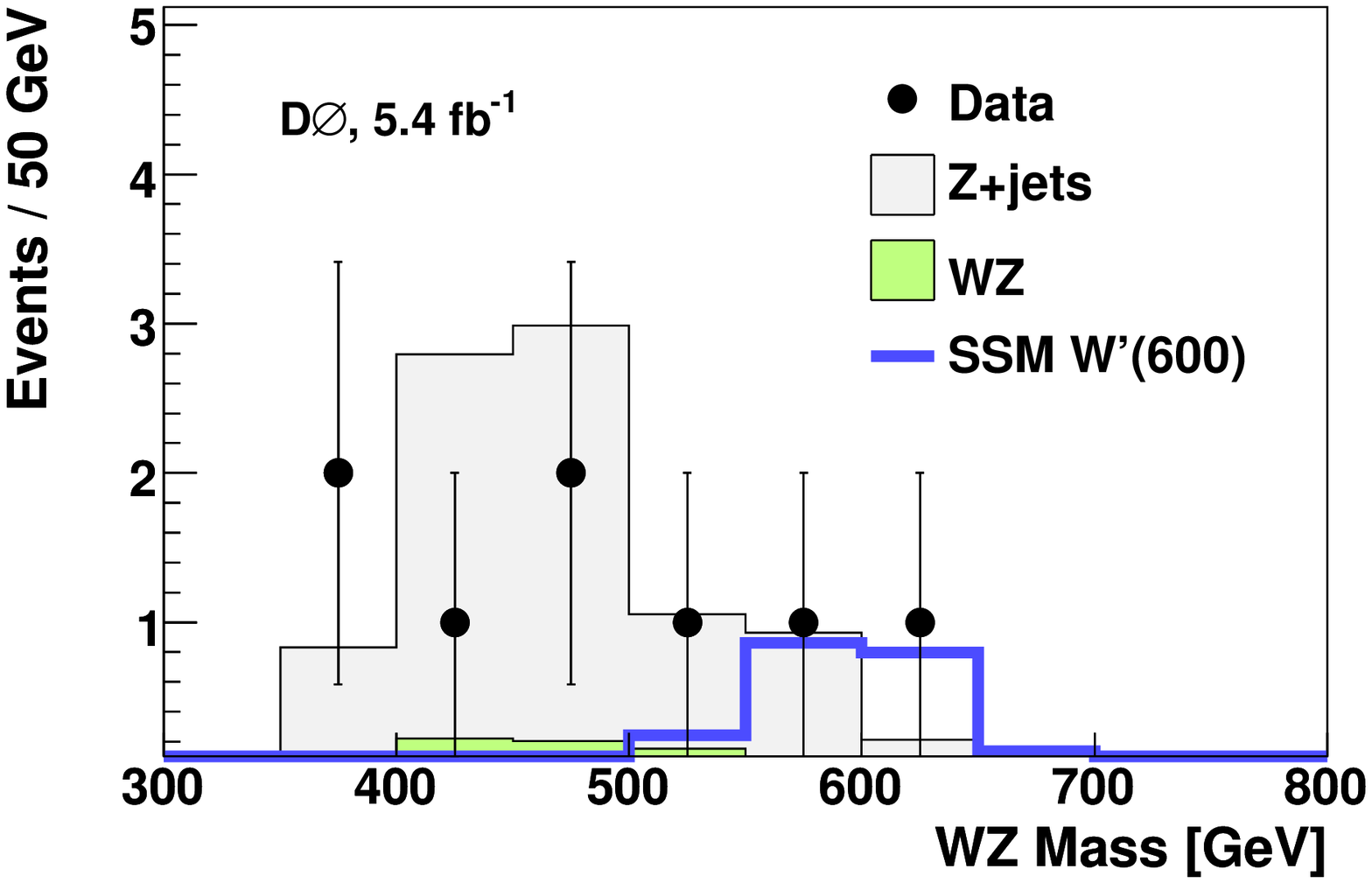}
    }}
   \put(35,43){\small {(a)}}
   \put(117,42){\small {(b)}}
  \end{picture}
  \caption{\label{fig:d0-WW-WZ}
   Distributions of the reconstructed $WW$ or $WZ$ transverse mass (left) and $WZ$ mass (right) for data (dots), estimated backgrounds,
and the estimated contributions of a 600~GeV SSM $W'$ and a 600~GeV $G$.}
\end{figure}

Production of pairs of gauge vector bosons has been observed
by CDF and D0 at the level predicted by the SM.
This sector is however not well constrained due to the smallness of the cross sections.
There is still room for an extra source of dibosons from the decay of massive charged
 or neutral particle as new gauge boson $W'$ or RS graviton $G$.
This has been looked at by CDF~\cite{cdf-dibosons}, and recently by D0~\cite{d0-dibosons}
using 5.4~fb$^{-1}$ of data.
One of the bosons is allowed to decay leptonically and the other hadronically leading to two event signatures ( {\it ll} + jets, {\it l} + jets + \MET)
where the lepton {\it l} is either an electron or a muon and \MET is the missing transverse energy carried away by the neutrino.
The two leptons or the lepton and \MET are first combined to form a $Z$ or $W$ candidate.
Since for a very high massive resonance, the two bosons would be highly boosted, the two hadronic showers
from the decay of the second boson may be reconstructed as a single massive jet.
So, D0 has increased its signal sensitivity by first trying to assign the $W$($Z$) hypothesis to
a jet with a mass larger than 60(70)~GeV before any two jet combination.
Figure~\ref{fig:d0-WW-WZ} shows for the data and the expected backgrounds the distributions of the reconstructed resonance 
and transverse masses in the dilepton and single lepton channels, respectively.
The predicted distributions of a 600~GeV $W'$ or $G$ are also shown.
Cross section limits are extracted from the absence of any significant excess of events in data from which one deduces lower bounds
 for $W'$ and $G$ masses of 690 and 754~GeV respectively.

\section{Search for vectorlike quarks}
\begin{figure}[htb]
  \centering
  \begin{picture}(160,55)
  \put(0,0){\hbox{
 \includegraphics[width=8.cm]{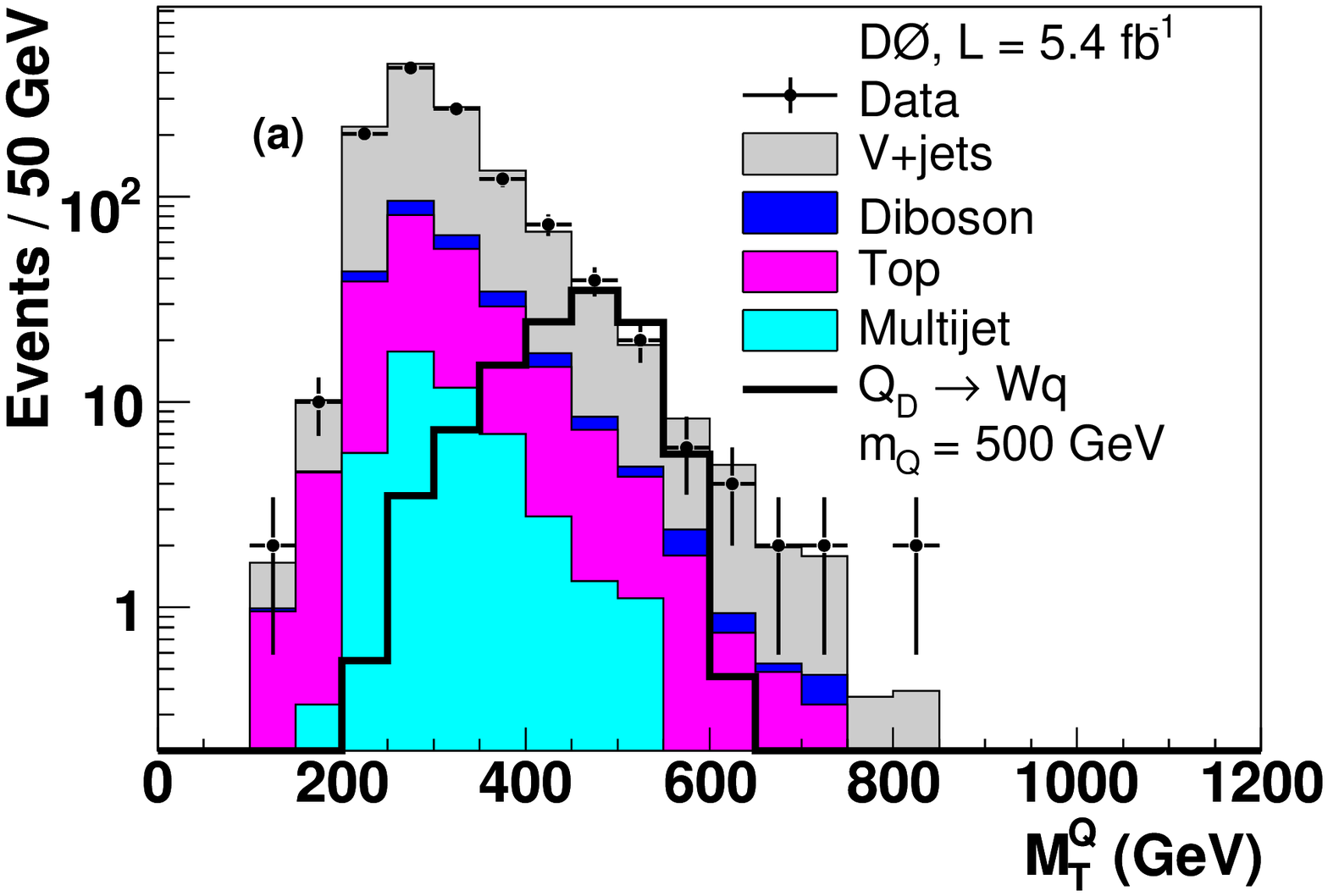}
 \includegraphics[width=8.cm]{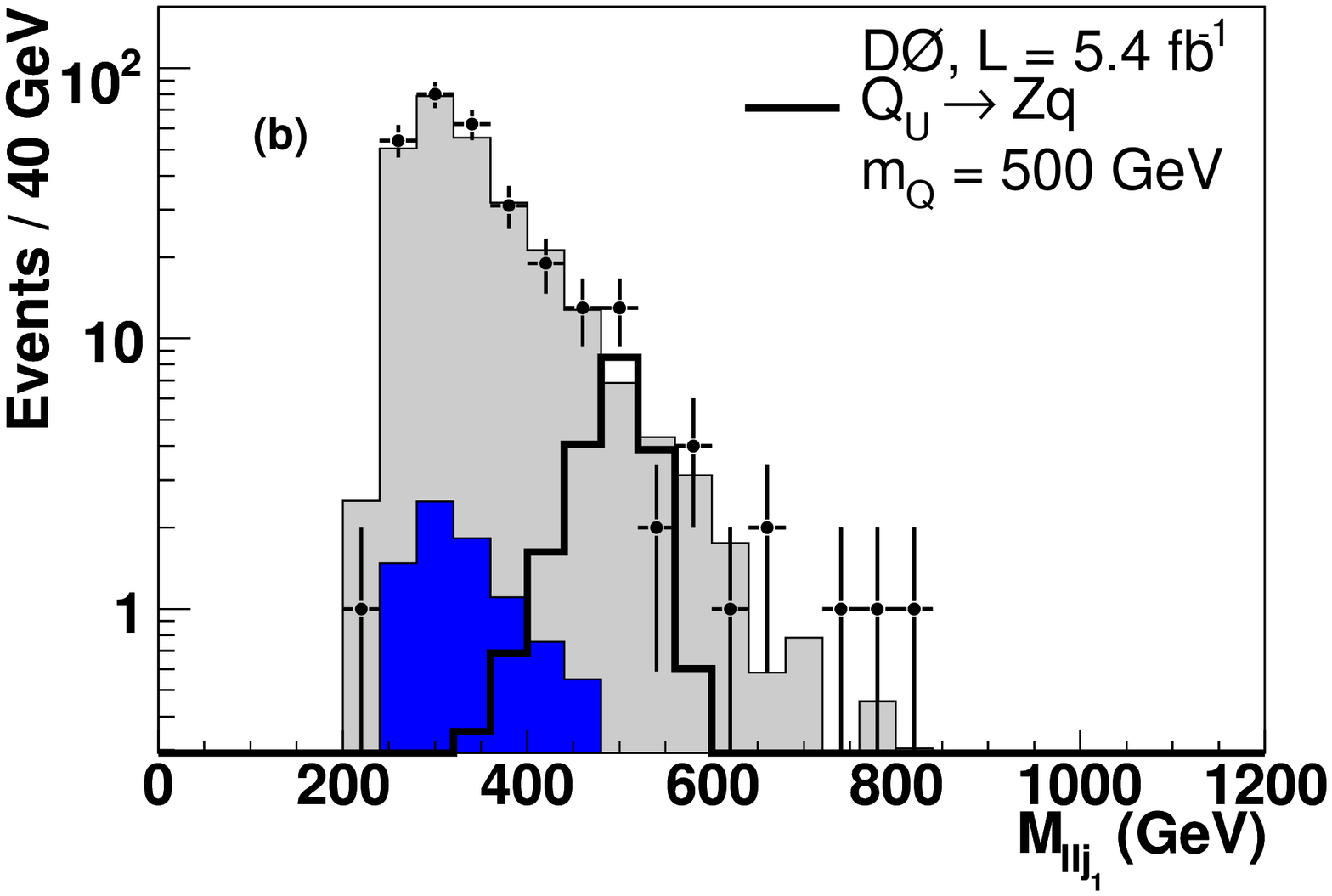}
    }}
  \end{picture}
  \caption{\label{fig:d0-VQ}
 (a) Vector-like quark transverse mass and (b) vector-like quark mass
for the single lepton and dilepton channels, respectively.
 Expected distributions for 500~GeV signals decaying as $Q_{D}\rightarrow Wq$ and $Q_{U}\rightarrow Zq$.
}
\end{figure}

Vectorlike quarks (VQ) share many characteristics of the SM quarks with the distinctive exception that their left and right components
transform in the same way under $SU(3)\times SU(2)\times U(1)$.
They can be singly produced via the electroweak interaction and may decay into a $W$ or $Z$ boson and a SM quark.
D0~\cite{d0-vector-quarks} has separated the analysis in two independant channels according to the leptonic decay of the vector
 boson ({\it ll} + jet, {\it l} +jet + \METnoSpace).
The event's leading jet in transverse momentum is assumed to come from the VQ decay.
Figure~\ref{fig:d0-VQ} shows, for the data and the expected backgrounds, the distributions of the reconstructed resonance 
and transverse masses in the dilepton and single lepton channels, respectively.
The absence of any statistically significant excess in data allows to derive cross section limits which are compared
to VQ production in two scenarios.
VQ not coupled to down-quarks and decaying exclusively to $Wq$ are excluded for masses below 693~GeV; if decaying to $Zq$
the lower bound of the mass is 551~GeV.
In the alternate scenario, i.e. no coupling to up-quarks, the mass limits are 403~GeV ($Wq$) and 430~GeV ($Zq$).

\vspace{-5mm}
\section{Search for new fermions (``Quirks'')}
A minimal extension of the standard model is obtained by the addition of a new unbroken SU(N) gauge group.
Such a group is characterized by the mass of the new fermions (quirks),Q, and the strength of the gauge coupling, $\Lambda$.
D0~\cite{d0-quirks} has considered the case where the quirks are charged, and $\Lambda << M_Q$, and $M_Q=0.1-1$~TeV.
 The breaking of the infra color is thus suppressed and a $Q\bar{Q}$ pair produced in $p\bar{p}$ collisions will not hadronize.
The  quirks in the pair will stay connected, as with a rubber band, the two tracks will not be resolved by the tracking system
 and they will be reconstructed as a single straight highly ionizing track.
This is the first search of this kind.
Event selection requires an isolated track and  a very high transverse momentum jet back to back. 
Such events are triggered by requiring jets and substantial \METnoSpace.
Analysing 2.4~fb$^{-1}$ of data, no excess of events at large ionization loss is observed over the expected background determined
from isolated tracks in an orthogonal data sample.
From the cross section limits  on the quirk production, D0 extract limits on the quirk mass depending on N, the number of colors
in the new gauge group, of 107, 119 and 133~GeV for N=2, 3 and 5 respectively.

\section{Search for leptonic jets}
\begin{figure}[htb]
  \centering
  \begin{picture}(160,55)
  \put(0,0){\hbox{
 \includegraphics[width=8.cm]{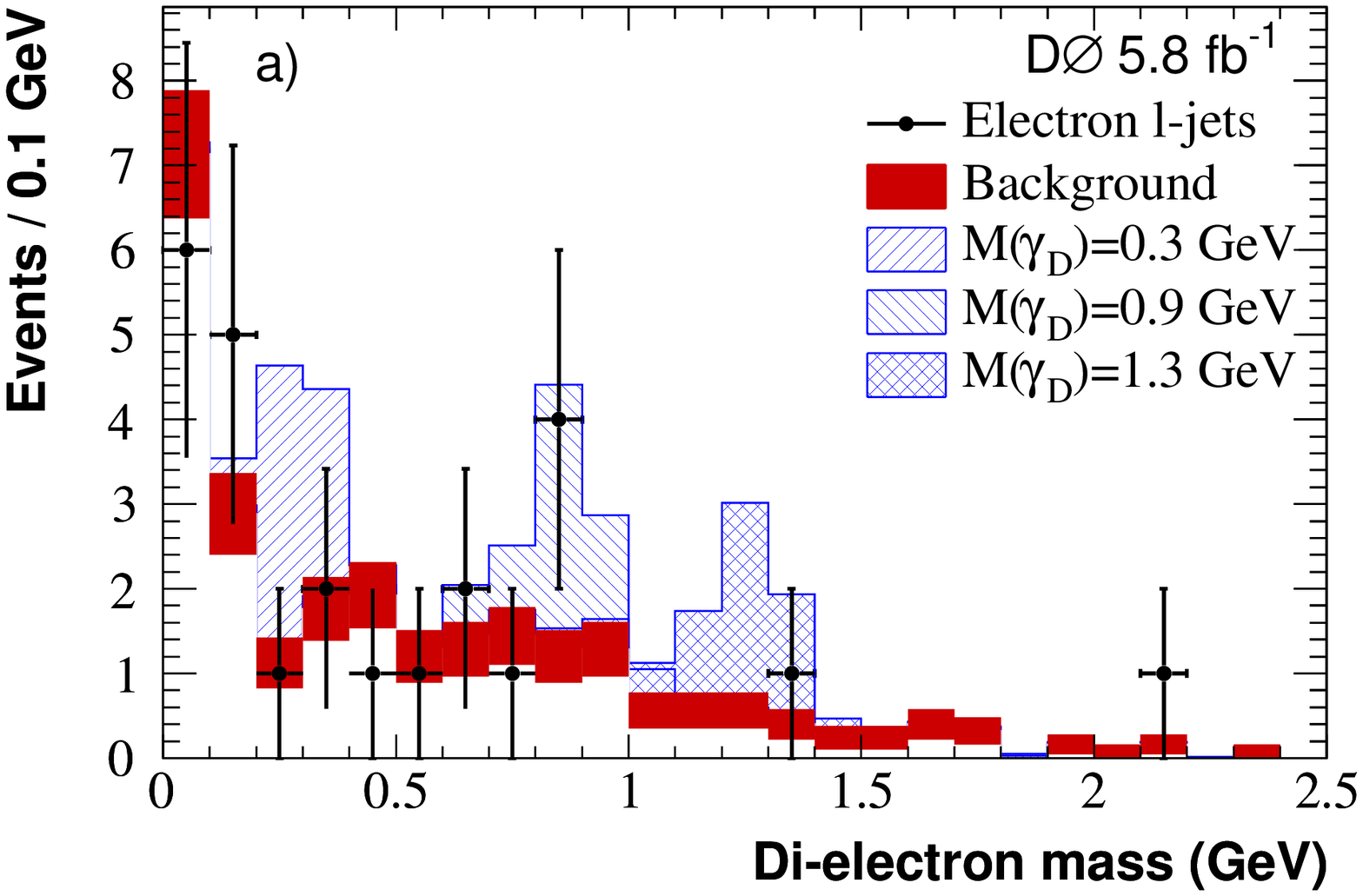}
 \includegraphics[width=8.cm]{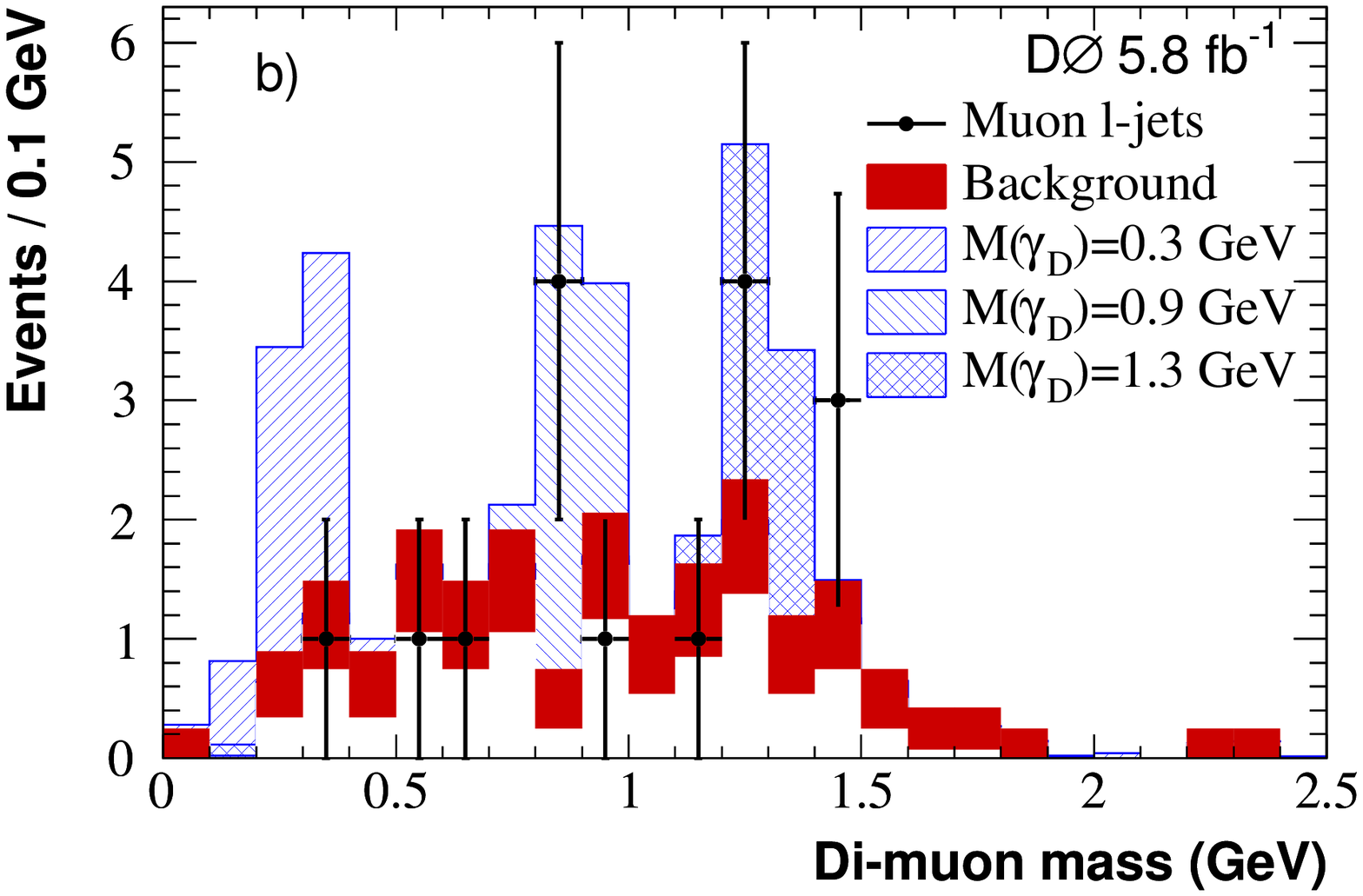}
    }}
  \end{picture}
  \caption{\label{fig:d0-ljet-masses}
Invariant mass of dark photon candidates with two isolated $l$-jets and $\MET>30$~GeV, for (a) electron
$l$-jets and (b) muon $l$-jets ( two entries per event, the $e\mu$ events contribute to both plots).
 The red band shows the shape of the mass distribution for events with $\MET<20$~GeV.
 The shaded blue histograms show the shapes of 8 MC signal events added to backgrounds, for three masses of the dark photon.
}
\end{figure}

Hidden Valley (HV) scenarios~\cite{HV} contain a hidden sector which is weakly coupled to SM particles.
They become popular as they provide convincing interpretation of observed astrophysical anomalies and discrepancies
in dark matter search.
New low mass particles are introduced in the hidden sector, and the dark photon, which is the force carrier, would have a mass
 around one GeV or less and would decay into a fermion or pion pair.
The case of decays to lepton pair ( electron or muon ) is particularly attractive. 
SUSY is often included in HV models, one could have a situation where the lightest neutralino will decay to a dark photon and $\tilde{X}$,
the lightest SUSY particle of the hidden sector, which will escape detection, leading to large \METnoSpace.
As the dark photon is light, it will be highly boosted in the neutralino decay, and the two leptons will be close to each other.
Experimentally, one has to change the isolation criteria usually applied to identify leptons.
The presence of a track of opposite charge close to the lepton candidate will sign the so-called leptonic jet ($l$-jet).
Using 5.8~fb$^{-1}$ of data, D0~\cite{d0-leptonic-jets} has searched for pair production of $l$-jets in three configurations :
$ee$, $\mu\mu$ and $e\mu$.
No evidence of $l$-jets is observed in the distributions of the electron and muon pair masses (Fig.~\ref{fig:d0-ljet-masses}).
Limits on the production cross section, around 100~fb for a 1~GeV dark photon, are obtained.

\vspace{-5mm}
\section{Search for a fourth generation down-type quark}
\begin{figure}[htb]
  \centering
  \begin{minipage}[t]{0.47\linewidth}
    \includegraphics[width=7.7cm]{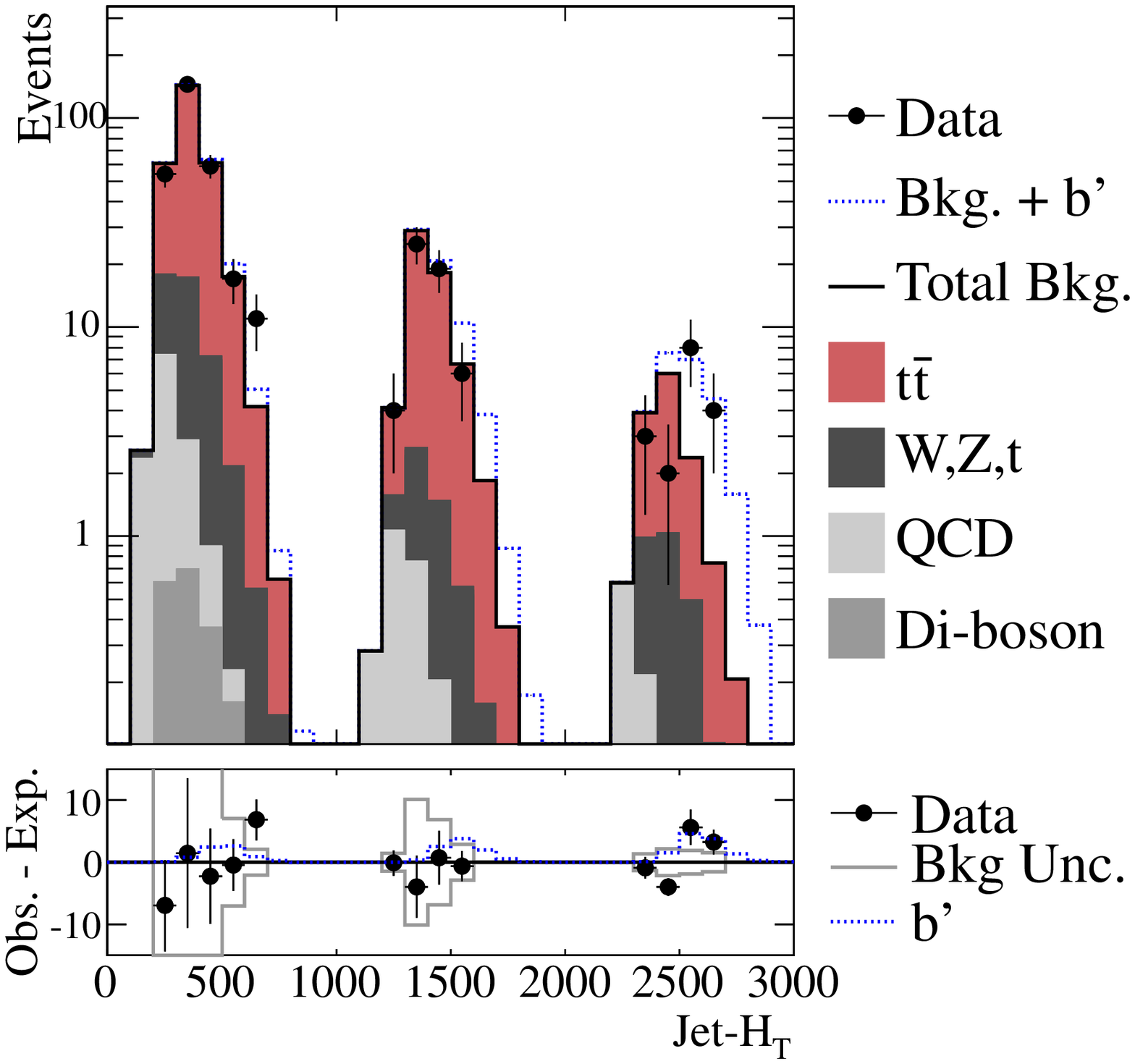}
    \caption{\label{fig:cdf-bprime-HT}
     Distribution of the variable Jet-$H_{T}$ for data and expected backgrounds.
     Contributions from a 350~GeV $b'$ signal on top of the expected backgrounds are also shown.
     The bottom pane shows the difference between the expected and observed number of events, as well as the total uncertainty on the expected number of events.
}
  \end{minipage}
  \hspace{5mm}
  \begin{minipage}[t]{0.47\linewidth}
    \includegraphics[width=7.7cm]{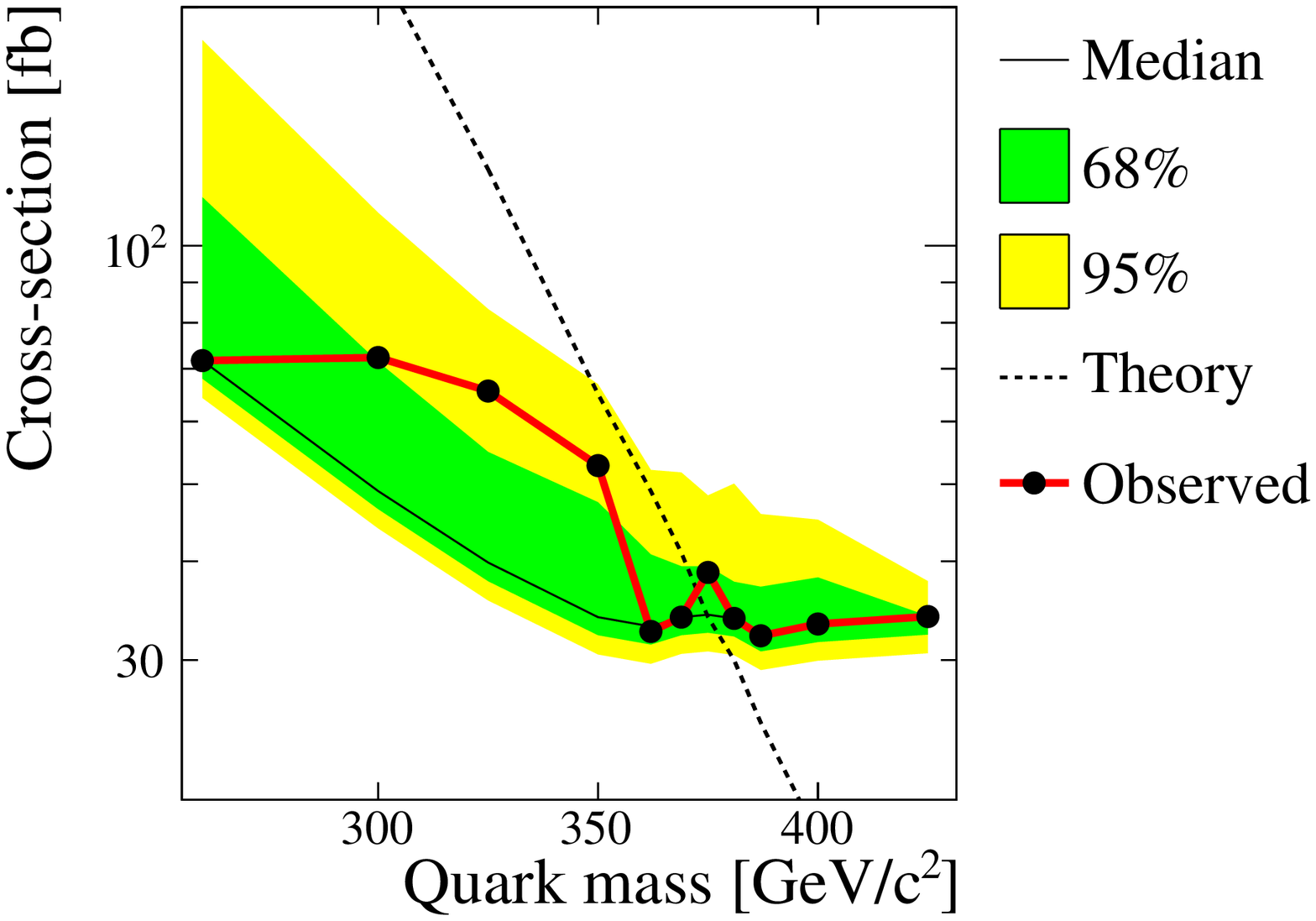}
    \caption{\label{fig:cdf-bprime-limit}
     95\% C. L. observed upper limits on $b'$ production cross section (red line) compared to the
     expected median limit (black line) in simulated experiments without $b'$ signal.
      Green and yellow bands represent 68\% and 95\%  of simulated experiments, respectively;
     the dashed line is the NLO $b'$ production cross section~\protect\cite{bprime-nlo}.
}
  \end{minipage}
\end{figure}

CDF~\cite{cdf-bprime} has searched for a fourth generation bottom-like quark ($b'$).
Current limits push $b'$ to be heavier than the sum of the $t$ quark and the charged gauge boson $W$ masses.
The analysis is realised using 4.8~fb$^{-1}$ of data assuming $b'$ to decay exclusively to $t$ and $W$.
The $b'$ pair produced in $p\bar{p}$ interactions will decay into two $b$ quarks and four $W$'s.
One of the $W$ decays leptonically and the others hadronically.
The final state is then characterised by an isolated lepton (electron or muon), \MET and many jets; one of the jets
is required to be tagged as a $b$-jet.
All the quark jets will not be reconstructed either because they fall below the transverse momentum cut or because their
hadronic showers are overlapping and they are merged in a single jet.
The analysis is performed in three jet multiplicity bins : 5, 6 and 7 jets or more.
The other variable which helps fighting the SM backgrounds, mainly $t\bar{t}$ and $W$ + jets production, is $H_T$,
the scalar sum of the transverse momentum of the lepton, jets and \METnoSpace.
Signal is expected to appear in the last multiplicity bin and for high $H_T$ values.
The $H_T$ distributions for the three multiplicity bins are shown in a single plot (Fig.\ref{fig:cdf-bprime-HT}) using the
variable $Jet-H_T$ equal to $H_T$, $H_T~+~1000~GeV$ or $H_T~+~2000~GeV$ for events with 5, 6, and 7 jets or more respectively.  
In the absence of any significant excess of events, cross section limits are set as a function of the $b'$ mass.
When compared to the NLO $b'$ cross section~\cite{bprime-nlo}, one gets  a lower bound for its mass of 372~GeV (Fig.~\ref{fig:cdf-bprime-limit}).

\section{Search for dark matter}
\begin{figure}[htb]
  \centering 
  \begin{picture}(160,65)
  \put(0,0){\hbox{
 \includegraphics[width=16.0cm]{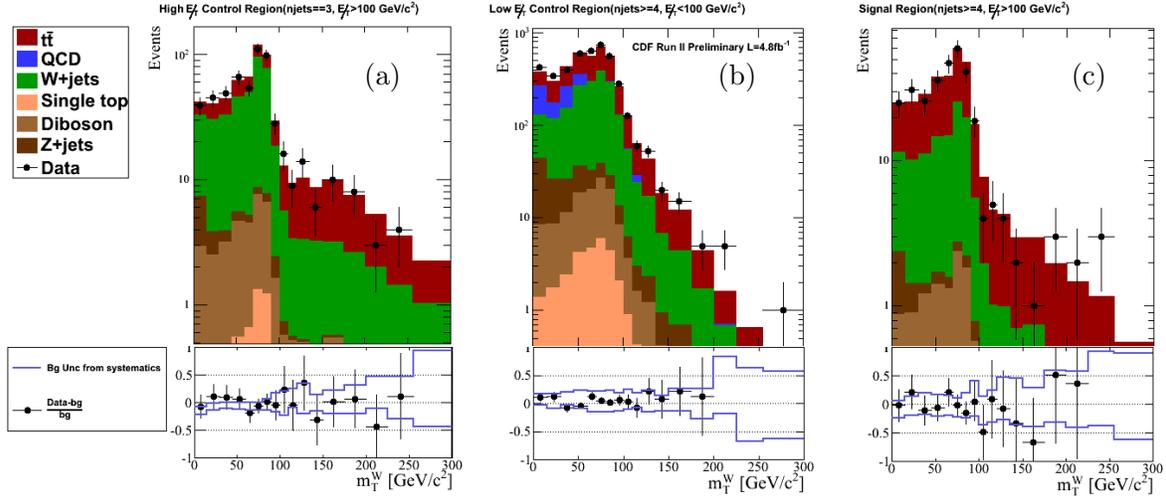}
    }}
   \put(50,55){(a)}
   \put(97,55){(b)}
   \put(144,55){(c)}
  \end{picture}
  \caption{\label{fig:cdf-darkmatter}
   Transverse mass distributions of the lepton and \MET for data and the expected backgrounds in control regions where signal is negligible (a and b) and where the $W$ is visible
and the region where it should appear (c).
   The bottom panes show the difference between expected background and observed events, as well as the total uncertainty on the expected
   background events.
}
\end{figure}

CDF~\cite{cdf-darkmatter} has searched for dark matter through the production of an exotic fourth generation $t'$ quark decaying to a $t$ quark and a dark matter
particle $X$.
The decay of the $t'$ to a $b$ quark and a charged gauge boson $W$ is not allowed.
The signal is searched for in the following event topology: an isolated lepton (electron or muon ), four jets or more, and very large \MET
due to the invisible particle $X$. 
The minimum value of \MET is optimised for each point in the ($t$' mass, $X$ mass) plane.
The larger the mass difference, the higher the cut.
The main SM backgrounds are $t\bar{t}$ pair and $W$ + jets production.
There are $W$ bosons in the main backgrounds which are prominently visible in the transverse mass distribution of the lepton and
\MET pair in the signal depleted domains i.e. at low \MET or low jet multiplicity as seen in Fig.~\ref{fig:cdf-darkmatter}a
and \ref{fig:cdf-darkmatter}b.
In the signal region, no significant excess of events is observed (Fig.~\ref{fig:cdf-darkmatter}c).
Cross section limits are obtained which allow to exclude a $t'$ mass below 360~GeV for a $X$ mass below 100~GeV.


\section{Conclusions}
The performance of the Tevatron has brought limits on BSM physics beyond one could have expected.
LHC experiments will take over, but CDF and D0 have still assets with their large datasets;
their analysis will be oriented towards complex final states.
Further details on physics results can be obtained from :
\begin{description}
\item[CDF] http://www-cdf.fnal.gov/physics/physics.html
\item[D0] http://www-d0.fnal.gov/Run2Physics/WWW/results.htm
\end{description}

\section*{Acknowledgments}
The author would like to thank the CDF and D0 working groups for providing
the material for this talk, and the organizers of the {\em Rencontres} for a
very enjoyable conference and the excellent organization.

\end{document}